\documentclass[a4paper,10pt]{article}

\usepackage{amsfonts, amsmath, amssymb}
\usepackage[utf8]{inputenc}
\usepackage[a4paper,top=3cm,bottom=2.5cm,left=2.5cm,right=2.5cm, bindingoffset=5mm]{geometry}
\usepackage{color}
\usepackage{booktabs}
\usepackage{subfigure}
\usepackage{graphicx}
\usepackage{float}
\usepackage{hyperref}
\hypersetup{citecolor=blue}
\usepackage{array}
\usepackage{colordvi}

\usepackage{mathtools}

\begin{document}

\vspace{20pt}

\begin{center}
{\bf{\Large Testing Lorentz invariance and CPT symmetry using gamma-ray burst neutrinos}}

\vspace{0.4cm} Xinyi Zhang$\mbox{}^{a}$,
Bo-Qiang Ma$\mbox{}^{a,b,c}$\footnote{Corresponding author, E-mail: \texttt{mabq@pku.edu.cn}}

\vspace{0.1cm} $\mbox{}^{a)}${ \em School of Physics and State Key Laboratory of Nuclear Physics and
Technology, \\}
{\em Peking University, Beijing 100871, China.}

\vspace{0.1cm} $\mbox{}^{b)}${\em Collaborative Innovation Center of Quantum Matter, Beijing, China}\\
\vspace{0.1cm} $\mbox{}^{c)}${\em Center for High Energy Physics, Peking University, Beijing 100871, China}
\end{center}

\vspace{.4cm}

\begin{abstract}
A recent work [Y. Huang and B.-Q. Ma, Commun. Phys. {\bf 1}, 62 (2018)] associated all four PeV neutrinos observed by IceCube to gamma-ray bursts (GRBs), and revealed a regularity which indicates a Lorentz violation scale $E_{\rm LV}=(6.5\pm0.4)\times10^{17}$ GeV with opposite sign factors $s=\pm 1$ between neutrinos and antineutrinos. 
The association of ``time delay" and
``time advance" events with neutrinos and antineutrinos (or vice
versa) is only a hypothesis since the IceCube detector cannot tell
the chirality of the neutrinos, and further
experimental tests are needed to verify this hypothesis.
We derive the values of the CPT-odd Lorentz violating parameters in the standard-model extension (SME) framework, and perform a threshold analysis on the electron-positron pair emission of the superluminal neutrinos (or antineutrinos). We find that different neutrino/antineutrino propagation properties, suggested by Y. Huang and B.-Q. Ma, can be described in the SME framework with both Lorentz invariance and CPT symmetry violation, but with a threshold energy constraint.  A viable way on testing the CPT symmetry violation between neutrinos and antineutrinos is suggested.
\end{abstract}

\vspace{.4cm}

Cosmic neutrinos from gamma-ray bursts (GRBs) are suggested to be ideal for studying the Lorentz invariance violation (LV)~\cite{Jacob:2006gn,AmelinoCamelia:2009pg,Amelino-Camelia:2015nqa}.
The IceCube Neutrino Observatory has observed four PeV neutrinos~\cite{Aartsen:2013bka,Aartsen:2014gkd}. The sources of such high-energy neutrinos are unknown until now, but widely believed to be extragalactic. The recent work in~\cite{Huang:2018ham} finds temporal and directional coincidence of these neutrinos with GRBs using a Lorentz-violation modified dispersion relation in an expanded time window. A regularity fitting well with these events is found, indicating a Lorentz-violation scale at $E_{\rm LV}=(6.5\pm0.4)\times10^{17}$ GeV, which is the same as that determined from IceCube events with energies ranging from 60 to 500~TeV~\cite{Amelino-Camelia:2016fuh,Amelino-Camelia:2016ohi}. More interestingly, both ``time delay'' and ``time advance'' events fit the regularity well, indicating that either neutrinos or antineutrinos are superluminal, while the other ones are subluminal~\cite{Huang:2018ham}. The association of ``time delay" and
``time advance" events with neutrinos and antineutrinos (or vice
versa) is only a hypothesis since the IceCube detector cannot tell
the chirality of the neutrinos, and further
experimental tests are needed to verify this hypothesis. The different propagation properties between neutrinos and antineutrinos can be explained by the CPT-odd feature of the linear Lorentz violation~\cite{Huang:2018ham}, indicating the charge, parity and time~(CPT) reversal symmetry violation between neutrinos and antineutrinos, or an asymmetry between matter and antimatter. In this work we discuss the implications of the findings in~\cite{Huang:2018ham} from a theoretical perspective. We also suggest a viable way on testing the CPT symmetry violation between neutrinos and antineutrinos.

We work in the standard-model extension (SME) framework~\cite{Colladay:1996iz,Colladay:1998fq}, which is an effective field theory including all operators of Lorentz violation with vast applications to study a large number of phenomena, such as modified dispersion relations of photons~\cite{Xiao:2009xe}
and fermions~\cite{Xiao:2008yu}, neutrino oscillation~\cite{Yang:2009ge,Diaz:2011ia} and neutrino superluminality~\cite{Qin:2011md,Ma:2011jj}, with parameters to be constrained by experimental observations. All the operators in the neutrino sector are classified and enumerated in Ref.~\cite{Kostelecky:2011gq}. The Lagrange density in the neutrino sector is~\cite{Kostelecky:2011gq}

\begin{eqnarray}
\mathcal{L}=\frac{1}{2}\bar{\Psi}_A(\gamma^\mu i \partial_\mu \delta_{AB}-M_{AB}+\hat{\mathcal{Q}}_{AB})\Psi_B+\mathrm{H.c}.,\label{eq:L}
\end{eqnarray}
where neutrinos and their charge conjugates are grouped in the multiplet $\Psi=(\nu_e,\nu_\mu,\nu_\tau,\nu_e^C,\nu_\mu^C,\nu_\tau^C)^T$, $M_{AB}$ is an arbitrary mass matrix, and $\hat{\mathcal{Q}}_{AB}$ is the Lorentz-violating operator. This operator can be decomposed in the basis of 16 Dirac matrices like

\begin{eqnarray}
\hat{\mathcal{Q}}_{AB}=\hat{\mathcal{S}}_{AB}+i\hat{\mathcal{P}}_{AB}\gamma_5+\hat{\mathcal{V}}_{AB}^\mu\gamma_\mu+\hat{\mathcal{A}}^\mu_{AB}\gamma_5\gamma_\mu+\frac{1}{2}\hat{\mathcal{T}}^{\mu\nu}_{AB}\sigma_{\mu\nu},
\end{eqnarray}
where each component is a $6\times6$ matrix that can be decomposed into $3\times3$ Dirac and Majorana blocks. An effective Hamiltonian can be constructed from the Lagrange density in Eq.(\ref{eq:L}).

For the discussion of the GRB neutrinos, it is safe to work in an oscillation-free model, in which the dispersion relation for a high-energy neutrino is~\cite{Kostelecky:2011gq}
\begin{eqnarray}
E(\boldsymbol{p})=|\boldsymbol{p}|+\sum_{djm}|\boldsymbol{p}|^{d-3}Y_{jm}(\boldsymbol{\hat{p}})[(a_{\rm of}^{(d)})_{jm}-(c_{\rm of}^{(d)})_{jm}],
\end{eqnarray}	
where $d$ is the effective mass dimension of the underlying operator, $j$, $m$ are angular-momentum indices, $(a_{\rm of}^{(d)})_{jm} (\text{where}~d\geq3, \text{odd})$ and $(c_{\rm of}^{(d)})_{jm} (\text{where}~d\geq4, \text{even})$ are oscillation-free coefficients for Lorentz violation. $(a_{\rm of}^{(d)})_{jm}$ associates with the CPT-odd operators while $(c_{\rm of}^{(d)})_{jm}$ are coefficients for the CPT-even operators. In this expression we use also the relativistic limit to omit the mass term.

It is natural to get the group velocity
\begin{eqnarray}
v_\nu=1+\sum_{djm}(d-3)|\boldsymbol{p}|^{d-4}Y_{jm}(\boldsymbol{\hat{p}})[(a_{\rm of}^{(d)})_{jm}-(c_{\rm of}^{(d)})_{jm}],\\
v_{\bar{\nu}}=1-\sum_{djm}(d-3)|\boldsymbol{p}|^{d-4}Y_{jm}(\boldsymbol{\hat{p}})[(a_{\rm of}^{(d)})_{jm}+(c_{\rm of}^{(d)})_{jm}].
\end{eqnarray}

Since the distribution of IceCube neutrino events is isotropic, we further simplify the expressions by assuming the rotation symmetry in the frame of the cosmic microwave background (CMB). Remember that the Earth-based frame has a boost velocity $\beta\simeq 10^{-3}$ compared to the CMB frame, an exact treatment should account for this factor cautiously. Here we take the isotropic limit as an illustration, the velocities are

\begin{eqnarray}
v_\nu=1+\sum_d(d-3)|\boldsymbol{p}|^{d-4}(\mathring{a}^{(d)}-\mathring{c}^{(d)}),\label{eq:v_sme1}\\
v_{\bar{\nu}}=1-\sum_d(d-3)|\boldsymbol{p}|^{d-4}(\mathring{a}^{(d)}+\mathring{c}^{(d)}),\label{eq:v_sme2}
\end{eqnarray}
where the isotropic coefficients $\mathring{a}^{(d)}=(a_{\rm of}^{(d)})_{00}/\sqrt{4\pi}$ and $\mathring{c}^{(d)}=(c_{\rm of}^{(d)})_{00}/\sqrt{4\pi}$.

In Ref.~\cite{Huang:2018ham}, using a general Lorentz-violation modified dispersion relation, the authors get the modified propagation velocity for neutrinos as
\begin{eqnarray}
v=1-s_n \frac{n+1}{2}(\frac{E}{E_{{\rm LV},n}})^n,\label{eq:v_lv}
\end{eqnarray}
where $n=1,2,...,$ corresponds to linear, quadratic, or higher order dependence of the energy, $s_n=\pm 1$ is a sign factor of Lorentz-violation correlation, and $E_{{\rm LV},n}$ is the $n$th-order Lorentz-violation scale. By taking into consideration only the linear energy dependence (i.e., $n=1$), the regularity observed by the authors indicates
\begin{eqnarray}
E_{\rm LV}=(6.5\pm0.4)\times10^{17} \mathrm{GeV},
\end{eqnarray}
which is close to the Planck scale $E_{\rm Pl}\simeq 1.22\times10^{19}$ GeV. It is worth noting that such a Lorentz-violation scale is compatible with that determined from
GRB photons~\cite{slj,zhangshu,xhw,xhw160509,Xu:2018ien,Liu:2018qrg} and it is also consistent with the constraints~\cite{Ellis:2018ogq,Wei:2018ajw} from recent coincident observation of a 290~TeV neutrino with the blazar TXS 0506+056~\cite{IceCube:2018dnn,IceCube:2018cha}. We emphasize here that the association of ``time delay" and ``time advance" events with neutrinos and antineutrinos (or vice versa) is only a hypothesis since the IceCube detector cannot tell the chirality of the neutrinos. As is stated in Ref.~\cite{Huang:2018ham} and is revealed in this paper, it is a reasonable one. Further experimental tests are needed to verify this hypothesis.

With the established correspondence of the velocity from the generalized Lorentz-violation modified dispersion relation in Eq.(\ref{eq:v_lv}) to the velocity from the isotropic SME model in Eqs.(\ref{eq:v_sme1}) and (\ref{eq:v_sme2}), and with also the constraint that neutrinos and antineutrinos have the same amounts of speed variation, we can relate $\mathring{a}^{(d)}$ to the Lorentz-violation scale $E_{{\rm LV},n}$. In the cases we keep only the leading term for a $d=5,7,9,11,\cdots$ (then $n=d-4=1, 3, 5, 7,\cdots$), we arrive at a general relation
\begin{eqnarray}
\mathring{a}^{(d)}\simeq\frac{1}{2 (E_{{\rm LV},d-4})^{d-4}},~\text{for superluminal}~~\nu;\label{eq:ad1}\\
\mathring{a}^{(d)}\simeq-\frac{1}{2 (E_{{\rm LV},d-4})^{d-4}},~\text{for superluminal}~~\bar{\nu}.\label{eq:ad2}
\end{eqnarray}

Since it is not clear whether neutrinos or antineutrinos are superluminal, we consider both possibilities. The $\mathring{a}^{(3)}$ term has no momentum dependence and only contribute as a constant energy shift, thus cannot be constrained by the velocity. Notice also the amounts of speed that departure from $1$, i.e., $\delta_v\equiv |v-1|$, are the same for both neutrinos and antineutrinos in Ref.~\cite{Huang:2018ham}, resulting in vanishing CPT-even coefficients $\mathring{c}^{(d)}$.

Apply these relations to the linear energy dependence case, which corresponds to the $\mathring{a}^{(5)}$ term, we immediately get
\begin{eqnarray}
\mathring{a}^{(5)}\simeq\frac{1}{2 E_{\rm LV}}\simeq 7.7\times10^{-19}~\text{GeV}^{-1},~\text{for superluminal}~~\nu;\\
\mathring{a}^{(5)}\simeq-\frac{1}{2 E_{\rm LV}}\simeq -7.7\times10^{-19}~\text{GeV}^{-1},~\text{for superluminal}~~\bar{\nu}.
\end{eqnarray}

There are still possibilities that the speed variation is caused by higher order energy dependence of Lorentz-violation terms when they are the leading term (lower order terms all vanish). A complete treatment would require a new fit to find $E_{{\rm LV},n}$ for a certain $n$ at the leading order, which is beyond the scope of the current paper. However, we can still get a sense of the situation by noticing that the observed time difference $\Delta t_{\rm obs}$, the redshift $z$ and the intrinsic time difference $\Delta t_{\rm in}$ stay the same for a certain association of a neutrino event with a GRB event. Since $\Delta t_{\rm obs}=\Delta t_{\rm LV}+(1+z)\Delta t_{\rm in}$, the LV time correction
\begin{eqnarray}
\Delta t_{\rm LV}=s_n\frac{1+n}{2 H_0}\frac{E_{\rm h}^n-E_{\rm l}^n}{E_{{\rm LV},n}^n}\int_0^z\frac{(1+z')^ndz'}{\sqrt{\Omega_m(1+z')^3+\Omega_\Lambda}}
\end{eqnarray}
stays the same for different $n$. Introducing an integral
\begin{eqnarray}
D_n(z)=\int_0^z\frac{(1+z')^ndz'}{\sqrt{\Omega_m(1+z')^3+\Omega_\Lambda}}
\end{eqnarray}
and the velocity variation
\begin{eqnarray}
\delta_{v_{n}}= \frac{n+1}{2}(\frac{E}{E_{{\rm LV},n}})^n,
\end{eqnarray}
we immediately get
\begin{eqnarray}
\delta_{v_{n}}D_n(z)= \delta_{v}D_1(z)=\frac{E}{E_{{\rm LV}}}D_1(z),
\end{eqnarray}
where the low energy $E_l$ is neglected compared to $E_h$.
Plugging in $\Omega_m=0.315$, $\Omega_\Lambda=0.685$, $E_{\rm LV}=6.5\times10^{17} \mathrm{GeV} $ with the highest energy event ($E_{\rm h}=2.6~\text{PeV}, z=2.15$) as an illustration, we can get $E_{{\rm LV},n}$. Then we find, e.g., the resulting CPT-odd coefficients $\mathring{a}^{(d)}$ from Eqs.(\ref{eq:ad1}) and (\ref{eq:ad2})
\begin{eqnarray}
\mathring{a}^{(7)}\simeq\pm1.2\times10^{-32}~\text{GeV}^{-3},\\
\mathring{a}^{(9)}\simeq\pm2.0\times10^{-46}~\text{GeV}^{-5},
\end{eqnarray}
where ``$+$" is the case neutrinos are superluminal, while ``$-$" corresponds to the case antineutrinos are superluminal. We list the results in Table~\ref{tab:a_lv}.

\begin{table}
 	\centering
 	 		\caption{Estimated values for the isotropic coefficients $\mathring{a}^{(d)}$ using the GRB neutrinos with {\it n}th-order energy dependence as the leading order ($n=d-4$). Units are GeV$^{4-d}$.}\label{tab:a_lv}
 		\begin{tabular}{l|c|c}
 			\toprule\midrule
 			& Superluminal $\nu$ & Superluminal $\bar{\nu}$ \\
 			\hline
 			$\mathring{a}^{(5)}$&$7.7\times10^{-19}$&$-7.7\times10^{-19}$\\
 			$\mathring{a}^{(7)}$&$1.2\times10^{-32}$&$-1.2\times10^{-32}$\\
 			$\mathring{a}^{(9)}$&$2.0\times10^{-46}$&$-2.0\times10^{-46}$\\
 			 			 \bottomrule
 		\end{tabular}
 		\end{table}

Superluminal neutrinos would lose energy through processes like, Cherenkov radiation ($\nu\rightarrow\nu\gamma$), neutrino splitting ($\nu \rightarrow\nu \nu\bar{\nu}$), electron-positron pair emission($\nu\rightarrow\nu e^+e^-$)~\cite{Cohen:2011hx}. Among these processes, the last one dominates the neutrino energy loss. High-energy neutrinos will lose energy until they are at or near the threshold energy. The same argument applies to antineutrinos for the CP-conjugated processes as understood. The threshold energy is estimated to be~\cite{Diaz:2013wia}

\begin{eqnarray}
E(\boldsymbol{p})&=&\sqrt{\boldsymbol{k}^2+m_e^2}+\sqrt{\boldsymbol{k}'^2+m_e^2}+E(\boldsymbol{p}')\nonumber\\
&\geq&\sqrt{(\boldsymbol{k}+\boldsymbol{k}')^2+4m_e^2}+\sqrt{\boldsymbol{p}'^2}\nonumber\\
&\geq&\sqrt{\boldsymbol{p}^2+4m_e^2}.
\end{eqnarray}
Squaring both sides and dropping the quadratic term in the Lorentz-violating part $\delta E(\boldsymbol{p})=E(\boldsymbol{p})-|\boldsymbol{p}|$, we arrive at

\begin{eqnarray}
|\boldsymbol{p}|\delta E(\boldsymbol{p})\simeq2m_e^2.
\end{eqnarray}

Assume only the CPT-odd terms, the observed neutrinos are near or below the threshold energy means

\begin{eqnarray}
	\sum_d|\boldsymbol{p}|^{d-2}\mathring{a}^{(d)}\leq2m_e^2,~\text{for superluminal}~~\nu;\label{eq:Eth1}\\
	-\sum_d|\boldsymbol{p}|^{d-2}\mathring{a}^{(d)}\leq2m_e^2,~\text{for superluminal}~~\bar{\nu}.\label{eq:Eth2}
\end{eqnarray}

We can calculate the limits on the coefficients $\mathring{a}^{(d)}$ from these expressions with neutrino (antineutrino) energy $2.6$ PeV (in order to make direct comparison with Table~\ref{tab:a_lv}). The results are shown in Table~\ref{tab:a_c} where the coefficients are taken to be nonzero one by one. We could also include the CPT-even terms, but it has been discussed in Ref.~\cite{Diaz:2013wia} where the authors use neutrino energy $2$ PeV. Interested readers should find results therein.

\begin{table}
	\centering
	\caption{Estimated bounds on the isotropic coefficients $\mathring{a}^{(d)}$ using the GRB neutrinos with threshold constraint from Cherenkov-like electron-positron emission. Units are GeV$^{4-d}$.}\label{tab:a_c}
	\begin{tabular}{l|c|c}
		\toprule\midrule
		& Superluminal $\nu$ ($\mathring{a}^{(d)}>0$)& Superluminal $\bar{\nu}$ ($\mathring{a}^{(d)}<0$)\\
		\hline
		$\mathring{a}^{(3)}$&$\le1.9\times10^{-13}$&$\ge-1.9\times10^{-13}$\\
		$\mathring{a}^{(5)}$&$\le2.8\times10^{-26}$&$\ge-2.8\times10^{-26}$\\
		$\mathring{a}^{(7)}$&$\le4.2\times10^{-39}$&$\ge-4.2\times10^{-39}$\\
		$\mathring{a}^{(9)}$&$\le6.2\times10^{-52}$&$\ge-6.2\times10^{-52}$\\
		\bottomrule
	\end{tabular}
\end{table}

The values we get in Table~\ref{tab:a_lv} do not satisfy the constraints set by the threshold effect in Table~\ref{tab:a_c}. To see this situation from another viewpoint, we find the threshold energy for the values of coefficients we get in Table~\ref{tab:a_lv} under the assumption that each term works at the leading order. We list the results in Table~\ref{tab:e_a}. We see that especially in the linear energy dependence case, the obtained $\mathring{a}^{(5)}$ value indicates a threshold energy of $8.7$ TeV. So the PeV neutrinos will go through the energy-loss process and will be depleted as a single emission causing a $78\%$ energy loss~\cite{Cohen:2011hx}. In other words, we will be unable to observe superluminal neutrinos (antineutrinos) of such energy.

From Table~\ref{tab:e_a}, we see that the threshold energy grows gradually with $d$. Given the condition that $\mathring{a}^{(13)}$ works at the leading order, the threshold energy will be around $1.2$ PeV, which is at the desired order. Although it is a peculiar situation in which all the lower $d$-terms vanish, it shows that the possibility of being compatible with the threshold limit does exist.

\begin{table}
	\centering
	\caption{Estimated threshold energy using the isotropic coefficients $\mathring{a}^{(d)}$ from Table~\ref{tab:a_lv}. List only the case that neutrinos are superluminal. Same threshold energy will be got for the case that antineutrinos are superluminal.}\label{tab:e_a}
	\begin{tabular}{l|c|c}
		\toprule\midrule
		&$\mathring{a}^{(d)}$ in GeV$^{4-d}$& Threshold Energy in GeV  \\
		\hline
		$\mathring{a}^{(5)}$&$7.7\times10^{-19}$&$8.7\times10^{3}$\\
		$\mathring{a}^{(7)}$&$1.2\times10^{-32}$&$1.3\times10^{5}$\\
		$\mathring{a}^{(9)}$&$2.0\times10^{-46}$&$4.2\times10^{5}$\\
		$\mathring{a}^{(11)}$&$3.2\times10^{-60}$&$8.1\times10^{5}$\\
		$\mathring{a}^{(13)}$&$5.0\times10^{-74}$&$1.2\times10^{6}$\\
		\bottomrule
	\end{tabular}
\end{table}

Keeping only the leading term and adopting $\mathring{a}^{(d)}$ from Eqs.(\ref{eq:ad1}) and (\ref{eq:ad2}) to the threshold constraints Eq.(\ref{eq:Eth1}), Eq.(\ref{eq:Eth2}), we have
\begin{eqnarray}
|\boldsymbol{p}|^{d-2}\leq4 m_e^2(E_{{\rm LV},d-4})^{d-4} ,~\text{for superluminal}~~\nu~~\text{or}~~\bar{\nu},
\end{eqnarray}
from which we arrive at the threshold (th) energy
\begin{eqnarray}
E_{\text{th}}\leq \sqrt[d-2]{4 m_e^2(E_{{\rm LV},d-4})^{d-4}} ,~\text{for superluminal}~~\nu~~\text{or}~~\bar{\nu},\label{eq:e_th}
\end{eqnarray}
which approaches to the corresponding Lorentz-violation scale $E_{{\rm LV},d-4}$ for enough large $d\to \infty$. For sufficient high $d$ at the leading order, the threshold energy $E_{\text{th}}$ will be sufficient high to ``protect" the superluminal high-energy neutrinos (antineutrinos). This relation also holds when $d$ is even.

For a superluminal GRB neutrino (antineutrino), the speed variation derived from a general Lorentz-violation modified dispersion relation is
\begin{eqnarray}
\delta_{v_n}=\frac{n+1}{2}(\frac{E}{E_{{\rm LV},n}})^n=\frac{d-3}{2}(\frac{E}{E_{{\rm LV},d-4}})^{d-4}\equiv\delta_{v_d},\label{eq:dv}
\end{eqnarray}
where $n=d-4$. Given the observed energy, time difference with the GRB photon, and the redshift of the GRB source, one can get $\delta_{v_d}$ for each superluminal event. Under the assumption that the observed highest neutrino energy is just around the threshold, i.e., $E\simeq E_{\text{th}}$, combine Eq.(\ref{eq:dv}) with Eq.(\ref{eq:e_th}), we get
\begin{eqnarray}
d\geq \frac{E^2}{2m_e^2}\delta_{v_d}+3,
\end{eqnarray}
which means, to get a compatible description with the threshold limit in the framework we adopt here, the leading order term should at least be $d$. In the current case, with $\delta_{v_d}\simeq (E/E_{\mathrm{LV}})D_1(z)/D_{d-4}(z)$ and $E\simeq 2.6$ PeV, we get $d\geq21$, which again, shows the difficulty in explaining the findings in Ref.~\cite{Huang:2018ham} in this framework while being compatible with the threshold limit.

It is worth mentioning that since only superluminal neutrinos (antineutrinos) go through these emission processes, the threshold effect only imposes limit on the superluminal part of data. Since only half of the data are superluminal in Ref.~\cite{Huang:2018ham}, it requires further comprehensive examination on this dataset. In case the superluminal part is invalid, e.g., the correlations of the neutrinos and GRBs are just coincident, the other half, being subluminal and rendering a same $E_{\rm LV}$ in the linear energy dependence case, can still be adjusted in the SME framework by performing a quadratic energy dependence fitting and values of the CPT-even coefficients $\mathring{c}^{(d)}$ will be got.

We propose to search for a pileup effect in neutrino (antineutrino) energy spectrum resulting from superluminal neutrinos (antineutrinos) that undergo Cherenkov-like emissions, as well as checking the speed variations of both superluminal and subluminal events that associated with GRBs, which, if are the same in both superluminal and subluminal cases, would be a strong sign of the CPT violation as indicated by the theory. It is also possible to observe the differences in superluminal neutrino (antineutrino) energy spectrum to distinguish the CPT-odd case with the CPT-even case~\cite{Stecker:2014oxa}.


Another thing worth noticing is that the regularity found in the neutrino sector exhibited in Ref.~\cite{Huang:2018ham} is similar to the light speed variation proposed previously from GRB photons~\cite{slj,zhangshu,xhw,xhw160509,Xu:2018ien,Liu:2018qrg}. To understand this similarity, new theoretical inputs beyond the SME are needed since in SME the coefficients in the two sectors are independent parameters to be constrained by fitting the experimental data.

To sum up, given the positive indication of Lorentz violation when associating the IceCube PeV neutrinos with GRBs, we describe the findings in Ref.~\cite{Huang:2018ham} in the SME framework. The existences of both ``time delay'' and ``time advance'' events with the same amounts of speed variations are well described by admitting only the CPT-odd terms of the theory, indicating a ``maximal'' CPT symmetry broken (``maximal'' in the sense that it requires only CPT-odd Lorentz violation terms). Our obtained values of the CPT-odd coefficients $\mathring{a}^{(d)}$ are below the limits in the data tables in Ref.~\cite{Kostelecky:2008ts} by 6$-$9 orders. However, the values of Lorentz-violation coefficients result in a lower energy threshold for the electron-positron pair emission process, which in turn means a constraint on the superluminal neutrino energy. We propose to further test both experimentally and theoretically the superluminal picture, investigate the possibility to resolve this paradox with new theoretical inputs, or find new mechanism coping with the pair emission process. 
The validity of the work in Ref.~\cite{Huang:2018ham} still needs to be checked and tested by more IceCube events in the future. An energy spectrum analysis in comparison with GRB models is also expected to testify the GRBs as the high-energy neutrino sources. We conclude that the new findings in Ref.~\cite{Huang:2018ham} can be described in the SME framework with both Lorentz invariance and CPT symmetry violation, but face challenge due to the constraint
on the superluminal neutrino energy from the threshold analysis. We need more evidence for the superluminal neutrinos from data, as well as novel insights to reconcile theories with observations.



\section*{Acknowledgements}

 This work is supported by National Natural Science Foundation of China (Grant No.~11475006).


\begin{thebibliography}{99}

	\bibitem{Jacob:2006gn}
	U.~Jacob and T.~Piran,
	Nature Phys.\  {\bf 3}, 87 (2007)
	[hep-ph/0607145].
	
\bibitem{AmelinoCamelia:2009pg}
  G.~Amelino-Camelia and L.~Smolin,
  Phys.\ Rev.\ D {\bf 80}, 084017 (2009)
  [arXiv:0906.3731 [astro-ph.HE]].

\bibitem{Amelino-Camelia:2015nqa}
  G.~Amelino-Camelia, D.~Guetta and T.~Piran,
  Astrophys.\ J.\  {\bf 806}, 269 (2015).



\bibitem{Aartsen:2013bka}
M.~G.~Aartsen {\it et al.} [IceCube Collaboration],
Phys.\ Rev.\ Lett.\  {\bf 111}, 021103 (2013)
[arXiv:1304.5356 [astro-ph.HE]].

\bibitem{Aartsen:2014gkd}
M.~G.~Aartsen {\it et al.} [IceCube Collaboration],
Phys.\ Rev.\ Lett.\  {\bf 113}, 101101 (2014)
[arXiv:1405.5303 [astro-ph.HE]].


\bibitem{Huang:2018ham}
Y.~Huang and B.-Q.~Ma,
Comms. Phys. {\bf 1}, 62 (2018)
[arXiv:1810.01652 [hep-ph]].


\bibitem{Amelino-Camelia:2016fuh}
	G.~Amelino-Camelia, L.~Barcaroli, G.~D'Amico, N.~Loret and G.~Rosati,
	Phys.\ Lett.\ B {\bf 761}, 318 (2016)
	[arXiv:1605.00496 [gr-qc]].
	
	
	\bibitem{Amelino-Camelia:2016ohi}
	G.~Amelino-Camelia, G.~D'Amico, G.~Rosati and N.~Loret,
	Nat.\ Astron.\  {\bf 1}, 0139 (2017)
	[arXiv:1612.02765 [astro-ph.HE]].
	
\bibitem{Colladay:1996iz}
D.~Colladay and V.~A.~Kostelecky,
Phys.\ Rev.\ D {\bf 55}, 6760 (1997)
[hep-ph/9703464].

\bibitem{Colladay:1998fq}
D.~Colladay and V.~A.~Kostelecky,
Phys.\ Rev.\ D {\bf 58}, 116002 (1998)
[hep-ph/9809521].

\bibitem{Xiao:2009xe}
  Z.~Xiao and B.-Q.~Ma,
 { Phys.\ Rev.\ D} {\bf 80}, 116005 (2009)
 [arXiv:0909.4927 [hep-ph]].

\bibitem{Xiao:2008yu}
  Z.~Xiao and B.-Q.~Ma,
  Int.\ J.\ Mod.\ Phys.\ A {\bf 24}, 1359 (2009)
  [arXiv:0805.2012 [hep-ph]].


\bibitem{Yang:2009ge}
  S.~Yang and B.-Q.~Ma,
  Int.\ J.\ Mod.\ Phys.\ A {\bf 24}, 5861 (2009)
  [arXiv:0910.0897 [hep-ph]].

 [arXiv:0911.0521 [hep-ph]].

\bibitem{Diaz:2011ia}
  J.~S.~Diaz and V.~A.~Kostelecky,
  Phys.\ Rev.\ D {\bf 85}, 016013 (2012)
  [arXiv:1108.1799 [hep-ph]].

\bibitem{Qin:2011md}
  N.~Qin and B.-Q.~Ma,
  Int.\ J.\ Mod.\ Phys.\ A {\bf 27}, 1250045 (2012)
  [arXiv:1110.4443 [hep-ph]].

\bibitem{Ma:2011jj}
  B.-Q.~Ma,
  Mod.\ Phys.\ Lett.\ A {\bf 27}, 1230005 (2012)
  [arXiv:1111.7050 [hep-ph]].


\bibitem{Kostelecky:2011gq}
V.~A.~Kostelecky and M.~Mewes,
Phys.\ Rev.\ D {\bf 85}, 096005 (2012)
[arXiv:1112.6395 [hep-ph]].





	\bibitem{slj}
	L.~Shao, Z.~Xiao and B.-Q.~Ma,
	{Astropart.\ Phys.\ } {\bf 33}, 312 (2010)
	[arXiv:0911.2276 [hep-ph]].

	\bibitem{zhangshu}
	S.~Zhang and B.-Q.~Ma,
	Astropart.\ Phys.\  {\bf 61}, 108 (2015)
    [arXiv:1406.4568 [hep-ph]].

	\bibitem{xhw}
	H.~Xu and B.-Q.~Ma,
	Astropart.\ Phys.\  {\bf 82}, 72 (2016)
	[arXiv:1607.03203 [hep-ph]].
	
	
	\bibitem{xhw160509}
	H.~Xu and B.-Q.~Ma,
	{Phys. Lett. B} {\bf 760}, 602 (2016)
    [arXiv:1607.08043 [hep-ph]].
	
\bibitem{Xu:2018ien}
  H.~Xu and B.-Q.~Ma,
  JCAP {\bf 1801},  050 (2018)
  [arXiv:1801.08084 [gr-qc]].

\bibitem{Liu:2018qrg}
  Y.~Liu and B.-Q.~Ma,
  Eur. Phys. J. C 78, 825 (2018)
  [arXiv:1810.00636 [astro-ph.HE]].

\bibitem{Ellis:2018ogq}
  J.~Ellis, N.~E.~Mavromatos, A.~S.~Sakharov and E.~K.~Sarkisyan-Grinbaum,
  Phys.\ Lett.\ B {\bf 789}, 352 (2019)
 [arXiv:1807.05155 [astro-ph.HE]].

\bibitem{Wei:2018ajw}
  J.~J.~Wei {\it et al.},
  arXiv:1807.06504 [astro-ph.HE].

	\bibitem{IceCube:2018dnn} 
	M.~G.~Aartsen {\it et al.} [IceCube and Fermi-LAT and MAGIC and AGILE and ASAS-SN and HAWC and H.E.S.S. and INTEGRAL and Kanata and Kiso and Kapteyn and Liverpool Telescope and Subaru and Swift NuSTAR and VERITAS and VLA/17B-403 Collaborations],
	Science {\bf 361}, no. 6398, eaat1378 (2018)
	[arXiv:1807.08816 [astro-ph.HE]].
	

	\bibitem{IceCube:2018cha} 
	M.~G.~Aartsen {\it et al.} [IceCube Collaboration],
	Science {\bf 361}, no. 6398, 147 (2018)
	[arXiv:1807.08794 [astro-ph.HE]].
	
\bibitem{Cohen:2011hx}
A.~G.~Cohen and S.~L.~Glashow,
Phys.\ Rev.\ Lett.\  {\bf 107}, 181803 (2011)
[arXiv:1109.6562 [hep-ph]].

\bibitem{Diaz:2013wia}
J.~S.~Diaz, V.~A.~Kostelecky and M.~Mewes,
Phys.\ Rev.\ D {\bf 89}, 043005 (2014)
[arXiv:1308.6344 [astro-ph.HE]].


	

\bibitem{Stecker:2014oxa}
F.~W.~Stecker, S.~T.~Scully, S.~Liberati and D.~Mattingly,
Phys.\ Rev.\ D {\bf 91}, no. 4, 045009 (2015)
[arXiv:1411.5889 [hep-ph]].

\bibitem{Kostelecky:2008ts}
V.~A.~Kostelecky and N.~Russell,
Rev.\ Mod.\ Phys.\  {\bf 83}, 11 (2011)
[arXiv:0801.0287 [hep-ph]].
	

\end{thebibliography}
\end{document}